\documentclass[10pt,aps,showpacs,showkeys,twocolumn,unsortedaddress,nofootinbib,floatfix]{revtex4-1}
\usepackage[T1]{fontenc}
\usepackage{amsmath}
\usepackage{times}
\usepackage{physics}
\usepackage{mathtools}
\usepackage{dsfont}
\usepackage{scalerel}
\usepackage[unicode=true, pdfusetitle, bookmarks=true, bookmarksnumbered=false, bookmarksopen=false, breaklinks=true, pdfborder={0 0 0}, backref=false, colorlinks=true, linkcolor=blue, citecolor=blue, urlcolor=blue]{hyperref}
\usepackage[dvipsnames]{xcolor}
\usepackage[normalem]{ulem}
\newcommand{\sm}{\kern0.1em}
\newcommand{\smalldiv}{\raisebox{-0.2ex}{\resizebox{!}{1.6ex}{\sm/\sm}}}
\newcommand{\tpsi}{\skew{3}{\tilde}{\psi}}
\makeatletter
\let\save@mathaccent\mathaccent
\newcommand*\if@single[3]{%
  \setbox0\hbox{${\mathaccent"0362{#1}}^H$}%
  \setbox2\hbox{${\mathaccent"0362{\kern0pt#1}}^H$}%
  \ifdim\ht0=\ht2 #3\else #2\fi
  }
\newcommand*\rel@kern[1]{\kern#1\dimexpr\macc@kerna}
\newcommand*\widebar[1]{\@ifnextchar^{{\wide@bar{#1}{0}}}{\wide@bar{#1}{1}}}
\newcommand*\wide@bar[2]{\if@single{#1}{\wide@bar@{#1}{#2}{1}}{\wide@bar@{#1}{#2}{2}}}
\newcommand*\wide@bar@[3]{%
  \begingroup
  \def\mathaccent##1##2{%
    \let\mathaccent\save@mathaccent
    \if#32 \let\macc@nucleus\first@char \fi
    \setbox\z@\hbox{$\macc@style{\macc@nucleus}_{}$}%
    \setbox\tw@\hbox{$\macc@style{\macc@nucleus}{}_{}$}%
    \dimen@\wd\tw@
    \advance\dimen@-\wd\z@
    \divide\dimen@ 3
    \@tempdima\wd\tw@
    \advance\@tempdima-\scriptspace
    \divide\@tempdima 10
    \advance\dimen@-\@tempdima
    \ifdim\dimen@>\z@ \dimen@0pt\fi
    \rel@kern{0.6}\kern-\dimen@
    \if#31
      \overline{\rel@kern{-0.6}\kern\dimen@\macc@nucleus\rel@kern{0.4}\kern\dimen@}%
      \advance\dimen@0.4\dimexpr\macc@kerna
      \let\final@kern#2%
      \ifdim\dimen@<\z@ \let\final@kern1\fi
      \if\final@kern1 \kern-\dimen@\fi
    \else
      \overline{\rel@kern{-0.6}\kern\dimen@#1}%
    \fi
  }%
  \macc@depth\@ne
  \let\math@bgroup\@empty \let\math@egroup\macc@set@skewchar
  \mathsurround\z@ \frozen@everymath{\mathgroup\macc@group\relax}%
  \macc@set@skewchar\relax
  \let\mathaccentV\macc@nested@a
  \if#31
    \macc@nested@a\relax111{#1}%
  \else
    \def\gobble@till@marker##1\endmarker{}%
    \futurelet\first@char\gobble@till@marker#1\endmarker
    \ifcat\noexpand\first@char A\else
      \def\first@char{}%
    \fi
    \macc@nested@a\relax111{\first@char}%
  \fi
  \endgroup
}
\makeatother

\begin{document}
\title{Questioning the adequacy of certain quantum arrival-time distributions}

\author{Siddhant Das}
\email{Siddhant.Das@physik.uni-muenchen.de}
\affiliation{Mathematisches Institut, Ludwig-Maximilians-Universit{\"a}t M\"{u}nchen, Theresienstr. 39, D-80333 M\"{u}nchen, Germany}
\author{Ward Struyve}
\email{ward.struyve@kuleuven.be}
\affiliation{Instituut voor Theoretische Fysica, KU Leuven, Belgium}
\affiliation{Centrum voor Logica en Filosofie van de Wetenschappen, KU Leuven, Belgium}


\begin{abstract}
It is shown that a class of exponentially decaying time-of-arrival probability distributions suggested by W\l{}odarz, Marchewka and Schuss, and Jurman and Nikoli{\'c}, as well as a semiclassical distribution implicit in time-of-flight momentum measurements, do not show the expected behavior for a Gaussian wave train. This casts doubts on the physical adequacy of these arrival-time proposals. In contrast, the quantum flux distribution (a special case of the Bohmian arrival-time distribution) displays the expected behavior.


\end{abstract}
%
%

\maketitle
\section{Introduction}
In view of the experimentally driven genesis of quantum mechanics and its notable empirical successes, it is a great surprise that a straightforward question such as ``\emph{how long} does it take for a quantum particle to strike the detector surface in a double-slit experiment?'' could be even more problematic than the question ``\emph{where}, in such an experiment, does the particle strike the detector surface?''. While the second question pertaining to the ubiquitous interference pattern is discussed in every quantum mechanics textbook and is experimentally well-established, the former concerning the arrival (or detection) time of the particle amenable to laboratory time-of-flight (TOF) experiments \cite{Pfau,*wig,*Pfaudetector} is a matter of an ongoing debate.

This seems almost paradoxical given that TOF measurements are the quintessence of methods determining, e.g., energies and momenta of particles \cite{COLTRIMS,Kothe,microscopy,fitting}, chemical reaction dynamics (as in the Rydberg tagging TOF technique \cite{RydbergTOF,RydbergTOF1}), or the temperature of single trapped atoms/ions \cite{RB,FerdiThermal}. However, it is not quantum mechanics that is invoked to interpret the TOF measurements in these cases. Instead, one employs various ansatzes and heuristics based on either Newtonian mechanics or geometric optics, whose capabilities for describing the data are highly questionable (especially in single-particle experiments featuring wave packet coherence).

That said, over the past decades, an increasing number of physicists have endeavored to formulate a first-principles description of arrival times within quantum mechanics, resulting in a multitude of disparate theoretical proposals for computing the arrival-time distribution\footnote{\(\Pi(\tau)\,d\kern-0.1em\tau\) is the probability that a particle prepared in a state \(\psi(x,0)\) at time zero is registered on a specified detector between time \(\tau\) and \(\tau+d\kern-0.1em\tau\).} \(\Pi(\tau)\) of a quantum particle \cite{MUGA1,MSP,MUGA}. However, experiments designed to help choose between competing viewpoints have been slow in coming.

The TOF distributions suggested in the literature can be divided into two broad categories. First, ideal (or intrinsic) arrival-time distributions that are \emph{apparatus-independent} theoretical predictions, given by some functional of the initial wave function \(\psi(x,0)\) and the geometrical surface of the detector (typically a single point on a line in one-dimensional discussions). A notable example is the quantum flux distribution
\begin{equation}\label{QFDist}
    \Pi_{\text{QF}}(\tau) =\frac{\hbar}{m}\,\text{Im}\big[\psi^*(L,\tau)\,\partial_x\psi(L,\tau)\big],
\end{equation} 
applicable for a particle of mass \(m\) arriving at the point \(x=L\) on a line. Here, \(\psi(x,t)\) denotes its wave function at time \(t\), a solution of Schr\"odinger's equation
\begin{equation}
    i\hbar\sm\frac{\partial\psi(x,t)}{\partial t}=-\,\frac{\hbar^2}{2m}\sm\frac{\partial^2\psi(x,t)}{\partial x^2}+V(x,t)\sm\psi(x,t),
\end{equation}
with initial condition \(\psi(x,0)\). The quantum flux distribution has been arrived at from various theoretical viewpoints, in particular, as the arrival-time distribution in Bohmian mechanics (de Broglie-Bohm or pilot-wave theory) \cite{DDGZ,*DDGZ96,Leav,*Leav98,Grubl,*Kreidl} in the absence of backflow \cite[p.\ 6]{Gaugeinv1}. Another well-known example applicable for freely moving particles, \(\smash{V(x,t)=0}\), is the Aharonov-Bohm \cite[Sec.\ 3]{MugaBadBad} and Kijowski \cite{kijTOF} arrival-time distribution \cite[Sec.\ 2]{Gaugeinv1}, which is typically indistinguishable from \(\Pi_{\text{QF}}(\tau)\) in the far-field or scattering regime accessible to present-day experiments.

Yet another ideal arrival-time distribution often implicit in TOF \emph{momentum} measurements is the semi-classical distribution 
\begin{equation}\label{Pisc}
    \Pi_{\text{SC}}(\tau) = \frac{m L}{\hbar\tau^2}\kern-0.1em\left|\tpsi\kern-0.15em\left(\frac{mL}{\hbar\tau}\right)\right|^2\!,
\end{equation}
where
\begin{equation}\label{Four}
    \tpsi(k) = \frac{1}{\sqrt{2\sm\pi}}\int_{-\infty}^{\infty}\!\!\!dx~\psi(x,0)\,e^{-\,ikx}
\end{equation}
is the Fourier transform of the wave function prepared at \emph{time zero} \cite{Hediffraction,NIST,microscopy,fitting}. This distribution is typically motivated along the following lines: For a classical trajectory \(\smash{x(t)=x(0)+pt\smalldiv m}\), with $x(0) \ll L$, the arrival-time of the particle is approximately given by \(\smash{\tau = m L\smalldiv p}\). The above distribution $\Pi_{\text{SC}}(\tau)$ is then obtained by considering this classical arrival-time formula, assuming that the width of \(\psi(x,0)\) is much smaller than \(L\) and that the momentum \(p\) is distributed according to the quantum mechanical momentum distribution \(\hbar^{-1}\sm |\tpsi(p\smalldiv\hbar)|^2\) \cite[p.\ 21]{NicolaPhd}. For a suitably localized \(\psi(x,0)\), the semiclassical distribution \eqref{Pisc} is also obtained from \(\Pi_{\text{QF}}(\tau)\) for a large \(L\) \emph{and} large \(\tau\) \cite{DDGZ,DDGZ96}.


The second category is that of non-ideal or measurement-inspired TOF distributions that involve a model of the detector. Various suggestions have been put forward, e.g., simple absorbing boundary conditions \cite{Werner,ABC}, complex potentials \cite{histories1,currentB,Allcock2}, wave function collapse (both detector induced \cite{zeno, JN} and spontaneous \cite{EEQT1,EEQT2}), path integrals with absorbing boundaries \cite{PI1,*PI2,*PI3}, a variety of quantum clocks \cite[Ch.\ 8]{MUGA}, and even a timeless formulation of QM \cite{Maccone}. An overview of these proposals, including a novel experimental set-up for distinguishing one from another (and in particular from \(\Pi_{\text{QF}}(\tau)\)) will appear in \cite{backwall1}.

In what follows, we focus on a class of non-ideal TOF distributions \cite{Wlodarz,PI1,*PI2,*PI3,JN} that have the form
\begin{equation}\label{PiL}
    \Pi(\tau) = \lambda(\tau)\sm\exp(-\!\int_0^{\tau}\!\!\!dt~\lambda(t)),
\end{equation}
where \(\lambda(t)\) is the so-called ``intensity function'' for which various proposals exist, see Table \ref{essenceTab}. This distribution is normalized as
\begin{equation}\label{normal}
    \int_0^{\infty}\!\!\!d\kern-0.1em\tau~\Pi(\tau)\,+\,P(\infty)=1,
\end{equation}
where 
\begin{equation}
  P(\infty)=\lim\limits_{\tau\to\infty}\frac{\Pi(\tau)}{\lambda(\tau)}
\end{equation}
is a ``non-detection probability'', accounting for the fraction of experimental runs in which the particle never arrives at \(L\). To derive \eqref{PiL} a time interval \([\sm0,\tau]\) is considered \cite[p.\ 2]{Wlodarz}, discretized into small time steps \(\smash{\Delta t=\tau\smalldiv N}\), where \(\lambda(t_n)\Delta t\) is the probability for the particle to be detected between time \(t_n=n\sm\Delta t\) and \(t_{n+1}=(n+1)\sm\Delta t\), \(n=0,\,1,\,2,\,\dots\sm N-1\). Further, assuming \emph{independent} probabilities at each time step\footnote{This generates an inhomogeneous Poisson point process with rate \(\lambda(t)\).}, the probability for the particle to be detected between time \(t_{N-1}\) and \(t_N\sm (=\tau)\) is simply
\begin{equation}\label{pn}
    \lambda(t_{N-1})\sm\Delta t\prod_{n=0}^{N-2}\big(1-\lambda(t_n)\sm\Delta t\big).
\end{equation}
By taking the limit \(\Delta t\to0\), equivalently \(N\to\infty\), of \eqref{pn}, the time-of-arrival density \eqref{PiL} is obtained. The intensity function $\lambda$ is supposed to follow from the physics of the detector.

\begin{table}[!ht]
\renewcommand{\arraystretch}{1.3}
\caption{Exponentially decaying arrival-time proposals and their intensity functions (the wave functions \(\overline{\psi}\) and \(\psi_c\) are defined in Sections \ref{Gausswavetrain} and \ref{ouranalysis}).}
\label{essenceTab}
\centering
\begin{tabular}{l c c}
    \hline
    \hline
    Proponents  & \(\smash{\lambda(t)}\) & Ref.\\
    \hline\\\\[-22pt]
    W\l{}odarz (W)   &   \(\displaystyle \smash{\lambda_0\sm|\psi(L,t)|^2}\) & \kern2mm\cite{Wlodarz}\\[15pt] Marchewka \& Schuss (MS) & \(\displaystyle \smash{(\lambda^\prime\epsilon\smalldiv\pi)\sm|\partial_x\sm\widebar \psi(L,t)|^2}\) & \kern2mm\cite{PI1,*PI2,*PI3}\\[15pt]
Jurman \& Nikoli\'c (JN)  &   \(\displaystyle \smash{\frac{1}{\delta t}\!\int_L^{L+\Delta L}\kern-0.5cm dx~|\psi_c(x,t)|^2}\) & \kern2mm\cite{JN}\\[15pt]
    \hline
    \hline
\end{tabular}
\end{table}
We will challenge these proposals by considering a train of Gaussian wave packets that initially have the same width and are moving with the same velocity towards the detector. By choosing the parameters so that each Gaussian wave packet reaches the detector one by one without significant spreading, it is expected on the basis of quasi-classical reasoning that each packet will contribute in the same way to the arrival-time distribution. In particular, it is expected that the arrival-time distribution will display peaks of the same shape and height at times roughly corresponding to the classically expected arrival times (i.e., the hitting times of individual packets). However, we will show that this is not the case for the proposals given in Table \ref{essenceTab}. While these distributions display peaks at the expected wave packet arrival times, the peaks are exponentially damped.

The outline of the paper is as follows: In Sec.\ \ref{Gausswavetrain} the Gaussian train is introduced. The analysis of the exponential proposals for this wave function follows next in Sec.\ \ref{ouranalysis}. The semiclassical distribution is treated in Sec.\ \ref{semi} and we conclude in Sec.\ \ref{discussion}.

\section{Gaussian wave train}\label{Gausswavetrain}
We direct our attention to the dynamics on a line, the detector occupying the interval \((L,L + \Delta L)\). Consider first a single Gaussian wave packet, initially ($t=0$) centered at \(x=0\), to the left of the detector, given by
\begin{equation}\label{chi0}
    \phi(x,0) = \frac{1}{\sqrt{\sigma\kern-0.1em\sqrt{\pi}}}\exp(-\,\frac{x^2}{2\,\sigma^2}+\frac{i}{\epsilon}\sm v\sm x).
\end{equation}
Here, \(\epsilon=\hbar\smalldiv m\), 
\begin{equation}\label{nainly}
    \sigma \ll L
\end{equation}
is the width of the wave packet, and \(v>0\) is the phase velocity. Under the free Schr\"odinger evolution, with Hamiltonian
\begin{equation}\label{freeHam}
    H=-\,\frac{\hbar^2}{2\sm m}\,\frac{d^2}{dx^2},
\end{equation}
the time-dependent packet is
\begin{align}\label{timedepsol}
    \phi(x,t)&= e^{-\,it\sm H/\hbar}\sm\phi(x,0)\nonumber\\
    &=\frac{1}{\sqrt{\sigma(t)\sqrt{\pi}}}\exp\kern-0.1em\left\{-\,\frac{\sigma}{\sigma(t)}\left[\frac{x^2}{2\,\sigma^2}-\,\frac{iv}{\epsilon}\!\left(x-\frac{vt}{2}\right)\right]\sm\right\}\!,
\end{align}
where
\begin{equation}\label{width}
    \sigma(t)=\sigma\left(1+i\frac{\epsilon t}{\sigma^2}\right)\!.
\end{equation}
The amplitude of this packet is
\begin{equation}\label{chit}
 |\phi(x,t)| = \frac{1}{\sqrt{|\sigma(t)|\sqrt{\pi}}}\exp(-\,\frac{1}{2}\sm\frac{(x-vt)^2}{|\sigma(t)|^2}).
\end{equation}
From this we see that the center of the packet arrives at the detector at time
\begin{equation}\label{hittingtime}
    \tau_0 = \frac{L}{v}.
\end{equation}
Hence, the corresponding time-of-arrival distribution $\Pi_0(\tau)$ is expected to be approximately peaked around $\tau_0$.

We also assume that the packet suffers negligible distortion during \(0<t<\tau_0\). This is guaranteed if
\begin{equation}\label{condition}
    \frac{\epsilon \tau_0}{\sigma^2}=\frac{\epsilon L}{\sigma^2 v}\ll1,
\end{equation}
which we shall refer to as the ``no spreading condition''. In this case,
\begin{equation}\label{mainly}
     |\phi(x,t)| \approx  |\phi(x-vt,0)|
\end{equation}
for \(\smash{0<t<\tau_0}\).

\begin{figure}[t]
    \centering
    \includegraphics[width=\columnwidth]{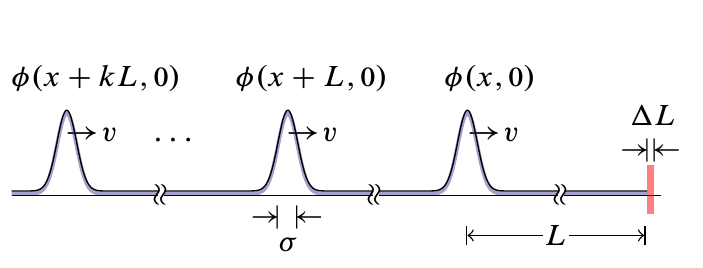}
    \caption{Train of Gaussian wave packets moving towards the detector.}
    \label{GT}
\end{figure}

Consider now an initial superposition of \(N\) Gaussian wave packets with the same width and velocity, but centered at \(\smash{x= -\, k L}\), \(\smash{k=0,\sm1,\sm\dots,\sm N-1}\), 
\begin{equation}\label{superpose}
    \psi(x,0) = \frac{1}{\sqrt{N}}\sum_{k=0}^{N-1}\phi(x+kL,0),
\end{equation}
as shown in Fig.\ \ref{GT}.
It evolves into
\begin{equation}\label{superposet}
    \psi(x,t) = e^{-\,it H/\hbar}\,\psi(x,0)=\frac{1}{\sqrt{N}}\sum_{k=0}^{N-1}\phi(x+kL,t),
\end{equation}
where \(\phi(\cdot,t)\) is given by \eqref{timedepsol}. Assuming
\begin{equation}\label{mainapprox}
    \smash{N\tau_0\ll\sigma^2\smalldiv\epsilon},
\end{equation}
the wave packets will remain non-overlapping until time \(N\tau_0\), the time at which the \(N\)\textsuperscript{th} wave packet of the Gaussian train strikes \(x=L\). As a consequence,
\begin{align}\label{distribute}
    |\psi(x,t)|^2 &\approx \frac{1}{N}\sum_{k=0}^{N-1}|\phi(x+kL,t)|^2\nonumber\\
                  &\overset{\eqref{mainly}}{\approx} \frac{1}{N}\sum_{k=0}^{N-1}|\phi(x+kL-vt,0)|^2
\end{align}
for all \(0<t<N\tau_0\). 

In this event, one expects the arrival-time distribution to have \(N\) peaks of nearly identical shape and height centered at times \(\smash{k\tau_0}\), \(k=1,\sm 2\sm\dots,\sm N\), i.e.,
\begin{equation}\label{expected}
    \Pi(\tau)\approx \frac{1}{N}\!\sum_{k=1}^{N}\Pi_0(\tau-k\tau_0).
\end{equation}
That is, the Gaussian wave packets in the train arrive at the detector one by one with time delays of \(\tau_0\), so that there should be a peak in the arrival-time density, which has the same shape for any Gaussian wave packet. 

The quantum flux TOF distribution \eqref{QFDist}, i.e.,
\begin{equation}\label{qfg}
    \Pi_{\text{QF}}(\tau)\approx \frac{v}{N\!\sqrt{\pi}\sm\sigma}\sum_{k=1}^{N}\exp[-\,\frac{v^2}{\sigma^2}\sm\big(\tau-k\tau_0\big)^2],
\end{equation}
which happens to agree with the Bohmian distribution in this case (due to the absence of backflow) {\em is} of the expected form \eqref{expected}. Note that the integral of \eqref{qfg} over all \(\tau>0\) is approximately unity, hence it predicts a zero non-detection probability as per Eq.\ \eqref{normal}.

\section{Exponential distributions}\label{ouranalysis}
The arrival-time distributions proposed in \cite{Wlodarz,PI1,*PI2,*PI3,JN} do not have the form \eqref{expected} for the Gaussian wave train. Instead, the probability density is exponentially falling off. In particular, assuming only \eqref{nainly} and \eqref{mainapprox}, we will show that the intensity functions (Table \ref{essenceTab}) take, for the Gaussian wave train \eqref{superpose}, the form
\begin{equation}\label{lambda}
    \lambda(t) \approx \frac{1}{N}\sum_{k=1}^N\lambda_0(t-k\tau_0),
\end{equation}
where $\lambda_0(t)$ is the intensity function corresponding to $\phi(x,t)$ supported on \(|t-\tau_0|\le \Delta\tau\), \(2\Delta \tau\) being the duration over which \(\phi\) sweeps over the detector \(\approx 3\sigma\smalldiv v\). This implies that the distribution \(\Pi(\tau)\) decays exponentially over time owing to the exponential factor in \eqref{PiL}. In fact, for \(\smash{k\tau_0<\tau<(k+1)\tau_0}\), we have
\begin{equation}
    \exp(-\!\int_0^{\tau}\!\!\!dt~\lambda(t)) \approx \exp(-\sm k\!\int_{\tau_{\scaleto{0}{3.5pt}}-\Delta\tau}^{\tau_{\scaleto{0}{3.5pt}}+\Delta\tau}\!\!\!\!\!dt~\lambda_0(t)).
\end{equation}
It follows that the expected behavior \eqref{expected} cannot hold.

\subsection{The W\l{}odarz proposal}
Using \eqref{distribute}, we readily obtain the intensity function $\lambda_{\text{W}}$ given in Table \ref{essenceTab}:
\begin{equation}\label{WWW}
    \lambda_{\text{W}}(t)\approx \frac{\lambda_0}{N}\sum_{k=1}^{N}|\phi(kL-vt,0)|^2,
\end{equation}
which implies the property \eqref{lambda}.

\subsection{The Marchewka \& Schuss proposal}
To calculate \(\lambda_{\text{MS}}(t)\), cf.\ Table \ref{essenceTab}, we need $\widebar \psi(x,t)$ which is the solution to Schr\"odinger's equation on the half-line $(-\infty,L]$ with initial condition \(\psi(x,0)\) and Dirichlet boundary condition at \(\smash{x=L}\). It is given by
\begin{equation}\label{varp}
    \widebar \psi(x,t)=\psi(x,t)-\psi(2\sm L-x,t)
\end{equation}
for \(x\le L\). (This state is only approximately normalized to unity, since \eqref{superpose} has support on $[L,\infty)$.) The (left) derivative at $L$ is \(\partial_x\sm \widebar \psi(L,t)=2\sm\partial_x\sm \psi(L,t)\) and
\begin{equation}\label{sumd}
    |\partial_x\sm \widebar \psi(L,t)|^2 \approx \frac{4}{N}\!\sum_{k=1}^N |\partial_x\phi(kL,t)|^2,
\end{equation}
where we used the fact that the $\phi(kL,t)$ and hence the $\partial_x\phi(kL,t)$ for different \(k\)s are approximately non-overlapping due to \eqref{mainapprox}.

To evaluate the summands we use \eqref{timedepsol}, obtaining
\begin{align}\label{exact}
    |\partial_x\phi(x,t)|^2 &= \left|\frac{\phi(x,t)}{\sigma(t)}\right|^2\left[\left(\frac{x}{\sigma}\right)^{\!2}+\left(\frac{v\sm\sigma}{\epsilon}\right)^{\!2}\right]\!,
\end{align}
which is exact, but in view of \eqref{mainapprox},
\begin{align}\label{approxi}
    |\partial_x\phi(kL,t)|^2\approx \left(\frac{v}{\epsilon}\right)^{\!2}\!|\phi(kL,t)|^2
\end{align}
for \(k=1,\sm2,\sm\dots\sm N\) and \(0<t<N\tau_0\). Then, using \eqref{mainly} and \eqref{sumd}, we arrive at
\begin{equation}
    \lambda_{\text{MS}}(t)\approx  \frac{4\sm v^2\lambda^\prime}{N\pi\epsilon}\sum_{k=1}^{N}\big|\phi(kL-vt,0)\big|^2.
\end{equation}
In this case, the intensity function approximately agrees with $\lambda_{\text{W}}$ (up to a proportionality factor). Again the property \eqref{lambda} obtains.

\subsection{The Jurman \& Nikoli\'c proposal}
To calculate \(\lambda_{\text{JN}}(t)\), given in Table \ref{essenceTab}, we need \(\psi_c(x,t)\), defined by
\begin{equation}\label{psic}
    \psi_c(x,t)=e^{-\,i\delta tH/\hbar}e^{-\,i(t-\delta t)\widebar{H}/\hbar}\,\psi(x,0),
\end{equation}
where \(H\) is the free Hamiltonian \eqref{freeHam} and $\widebar{H}$ refers to free motion on the half-line \((-\infty,L]\) with Dirichlet boundary conditions at \(L\). The initial wave function is assumed to be supported on the half-line. Our initial wave function \eqref{superpose} is actually nonzero in the region $[L,\infty)$ but, as before, we will ignore its tail beyond \(x=L\). So, using the notation introduced in \eqref{varp}, we can also write
\begin{equation}\label{psic2}
    \psi_c(x,t)=e^{-\,i\delta tH/\hbar}\,\widebar \psi(x,t-\delta t).
\end{equation}

To evaluate \eqref{psic} for the Gaussian train, consider for \(0\le k<N\), the function \(\phi_c\) defined by
\begin{align}
    \phi_c(x+kL,t)&\coloneqq e^{-\,i\delta tH/\hbar}e^{-\,i(t-\delta t)\widebar{H}/\hbar}\,\phi(x+kL,0) \label{defn}\\
    &=e^{-\,i\delta tH/\hbar}e^{-\,i(t-k\tau_{\scaleto{0}{3pt}}-\delta t)\widebar{H}/\hbar}\nonumber\\[-3pt]
    &\kern2cm\times\,e^{-\,ik\tau_{\scaleto{0}{3pt}}\widebar{H}/\hbar}\,\phi(x+kL,0)\nonumber\\
    &=e^{-\,i\delta tH/\hbar}e^{-\,i(t-k\tau_{\scaleto{0}{3pt}}-\delta t)\widebar{H}/\hbar}\, \widebar \phi(x+kL,k\tau_0) .\label{continue}
\end{align}
Since \(\phi(x+kL,k\tau_0)\) is centered at \(x=0\) and has a width \(|\sigma(k\tau_0)|\approx \sigma\) in view of the ``no spreading'' condition \eqref{mainapprox} [cf\ Eq.\ \eqref{width}], we have
\begin{equation*}
    \widebar \phi(x+kL,k\tau_0) \approx \phi(x+kL,k\tau_0). 
\end{equation*}

Using \eqref{mainly} and \eqref{mainapprox}, the amplitude of this wave function satisfies
\begin{equation}\label{nice}
    \big|\phi(x+kL,k\tau_0)\big| \approx |\phi(x,0)|.
\end{equation}
Its phase is \begin{align}
    \text{Arg}\sm\big[\phi(x+kL,k\tau_0)\big] &= \text{Arg}\sm\big[\phi(x,0)\big]-\frac{\epsilon k\tau_0}{2\,\sigma^2}  \nonumber\\
    &\sm+\,\frac{k\tau_0}{2}\kern-0.1em\left[\frac{v^2}{\epsilon}+\,\epsilon\,\frac{(x\smalldiv\sigma)^2}{|\sigma(k\tau_0)|^2}\right]\!.
\end{align}
Equation \eqref{mainapprox}, together with the condition \(\smash{|x|\lesssim 3\sm\sigma}\) valid within the bulk of the support of the wave function, allow us to neglect both the second term and the second term in brackets, hence 
\begin{equation*}
    \text{Arg}\sm\big[\phi(x+kL,k\tau_0)\big]\approx \text{Arg}\sm\big[\phi(x,0)\big]+k\sm\frac{v L}{2\sm\epsilon}.
\end{equation*}
It follows that 
\begin{equation*}
\phi(x+kL,k\tau_0)\approx \phi(x,0)\sm e^{i k v L/2\epsilon}
\end{equation*}
and
\begin{align}
    \phi_c(x+kL,t)&\approx e^{-\,i\delta tH/\hbar}e^{-\,i(t-k\tau_{\scaleto{0}{3pt}}-\delta t)\widebar{H}/\hbar}\phi(x,0)\sm e^{i k v L/2\epsilon} \nonumber\\
    &\overset{\eqref{defn}}{=}\sm e^{i k v L/2\epsilon}\phi_c(x,t-k\tau_0).
\end{align}
Hence, by linearity 
\begin{equation}
    \psi_c(x,t)\approx \frac{1}{\sqrt{N}}\sum_{k=0}^{N-1}e^{i k v L/2\epsilon}
    \phi_c(x,t-k\tau_0).
\end{equation}
Ignoring the tails of the Gaussians, we have that for $\delta t  \ll \tau_0 $, at most only one of the wave packets $\phi_c(x,t-k\tau_0)$ will have its support in $[L, L+\Delta]$ at a given time. (Remember that $\phi_c$ is obtained by free evolution with Dirichlet boundary conditions up until time $t-\delta t$, and then free evolution for a time $\delta t$.) Therefore we can ignore cross terms for the density in  the interval $[L, L+\Delta]$, and write
\begin{align}
    \lambda_{\text{JN}}(t) &\approx \frac{1}{\delta t}\sum_{k=0}^{N-1}\int_L^{L+\Delta L}\!\!\!\!\!\!\!dx~|\phi_c(x,t-k\tau_0)|^2,
\end{align}
so that property \eqref{lambda} obtains.

\section{The semiclassical distribution}\label{semi}
\begin{figure}[!ht]
    \centering
    \includegraphics[trim=0 1mm 3mm 3mm, clip, width=\columnwidth]{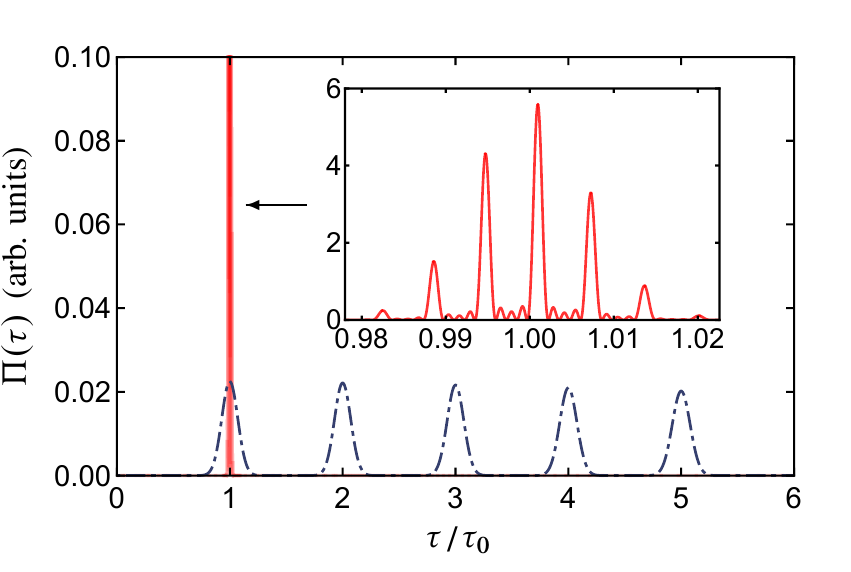}
    \caption{An illustration of \(\Pi_{\text{SC}}(\tau)\) and \(\Pi_{\text{QF}}(\tau)\) (dot-dashed) for parameter values \(N=5\), \(\sigma=5\), \(L=10\sm\sigma\), \(v=1\), and \(\epsilon=0.05\) (in arbitrary units). Inset: Magnified view of \(\Pi_{\text{SC}}(\tau)\).}
    \label{SCQF}
\end{figure}

The semiclassical distribution, cf.\ Eqs.\ (\ref{Pisc}-\ref{Four}), is
\begin{equation}\label{SC}
    \Pi_{\text{SC}}(\tau) = \frac{L}{\epsilon\tau^2}\kern-0.1em\left|\frac{1}{\sqrt{2\sm\pi}}\!\int_{-\infty}^{\infty}\!\!\!dx~\psi(x,0)\,e^{-\,iLx/\epsilon\tau}\right|^2\!.
\end{equation}
Using the Fourier transform
\begin{align}
    &\frac{1}{\sqrt{2\sm\pi}}\!\int_{-\infty}^{\infty}\!\!\!dx~\phi(x+x_0,0)\,e^{-\,iv_{\scaleto{0}{3pt}}\sm x/\epsilon} \nonumber\\
    &\qquad= \sqrt{\frac{\sigma}{\pi^{1/2}}}\,\exp[\frac{i}{\epsilon}\sm v_0 \sm x_0-\sm\frac{\sigma^2}{2\sm\epsilon^2}\left(v-v_0\right)^{\!2}],
\end{align}
we find (without approximations) that
\begin{align}
    \Pi_{\text{SC}}(\tau) &= \frac{\sigma L}{N\kern-0.1em\sqrt{\pi}\sm\epsilon\sm\tau^2}\,\frac{\sin^2(NL^2\!\smalldiv2\sm\epsilon\tau)}{\sin^2(L^2\!\smalldiv2\sm\epsilon\tau)}\nonumber\\
    &\kern2cm \times\exp[-\,\frac{\sigma^2v^2}{\epsilon^2}\left(1-\tau_0/\tau\right)^2].
\end{align}
The distribution is peaked around $\tau_0$, contrary to what is expected of the Gaussian wave train. This should come as no surprise since \(\Pi_{\text{SC}}\) is fully determined by the momentum distribution of the initial wave function, which in the present example is centered around \(\smash{p=mv}\). While the semiclassical distribution sometimes follows from the quantum flux/Bohmian distribution, such is not the case here, as can be seen in Fig.\ \ref{SCQF}. This explains why the latter \emph{does} show the expected behavior, unlike the former.
\section{Discussion and outlook}\label{discussion}
\begin{figure}
    \centering
    \includegraphics[trim=0 1mm 3mm 3mm, clip, width=\columnwidth]{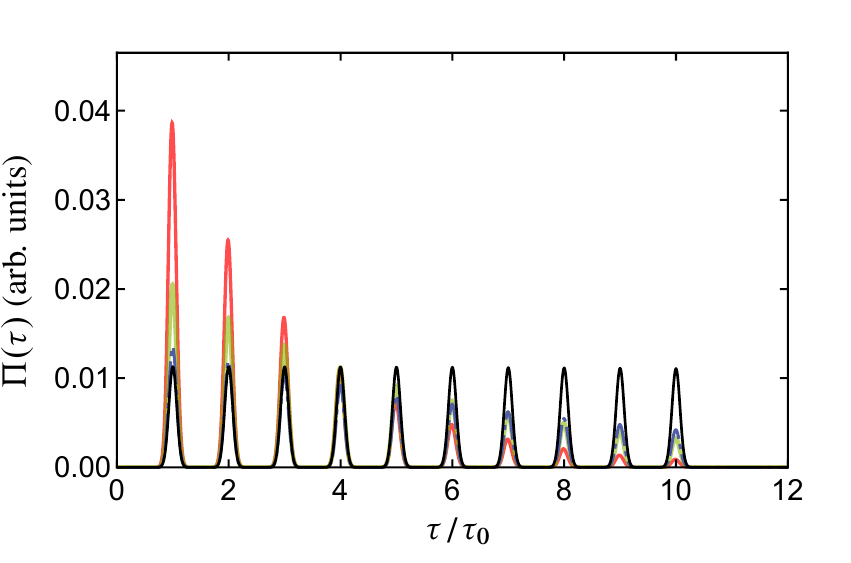}
    \caption{An illustration of the exponential decay of the arrival-time distributions for a Gaussian wave train satisfying the no spreading condition \eqref{mainapprox}, with parameters \(N=10\), \(\sigma=5\), \(L=10\sm\sigma\), \(\epsilon=0.01\), and \(v=1\) (in arbitrary units). The red curve is \(\Pi_{\text{JN}}\) with \(\delta t=0.5\) and \(\Delta L=0.5\sm\sigma\), the green curve is \(\Pi_{\text{W}}\) with \(\lambda_0=2\), while the blue curve denotes \(\Pi_{\text{MS}}\) with \(\lambda^\prime=0.01\). The black curve is \(\Pi_{\text{QF}}\), which does not display an exponential decay. The exponential TOF curves were calculated numerically without any approximations (see text for details).}
    \label{compare}
\end{figure}
The exponential distributions for the Gaussian train \eqref{superpose} are plotted in Fig.\ \ref{compare} along with the quantum flux/Bohmian distribution. All arrival-time distributions were produced using the exact analytic expressions without invoking approximations \eqref{nainly} and \eqref{mainapprox}\sm\footnote{For the Jurman \& Nikoli\'c proposal, we used 
$$\lambda_{\text{JN}}(t)\approx (\Delta L\smalldiv \delta t)\sm\big|\psi_c(L,t)\big|^2,$$
applicable for a small \(\Delta L\). However, \(\psi_c(x,t)\) [Eq.\ \eqref{psic2}] was evaluated exactly in terms of error functions by integrating \eqref{varp} against the known free-particle propagator for time \(\delta t\).}. While the quantum flux distribution displays the expected behavior (i.e., featuring identical and well-separated peaks centered at times \(\tau_0\), \(2\sm\tau_0\), \dots,\(\sm10\sm\tau_0\)), the exponential ones do not, thus substantiating our analysis. The free parameters \(\lambda_0\), \(\lambda^\prime\), \(\Delta L\), and \(\delta t\) were chosen for best visibility. However, this undesirable behavior cannot be evaded by tuning these parameters: making them larger causes a faster decay, while making them smaller moderates the decay at the cost of increasing the non-detection probability (to the extent of a vanishing arrival-time density in the case of an appreciable removal of the decay). In fact, given any choice of these free parameters, the number \(N\) of Gaussians in the train and their velocity \(v\) could be so chosen that the exponential decay practically washes out the arrival-time peaks corresponding to the trailing Gaussians.

The exponential proposals were aimed at deriving the TOF distribution by means of a detector model. While different intensity functions \(\lambda\) can be considered, our results show that the failure is not so much attributable to the particular choice of \(\lambda\) but presumably the assumption of \emph{independence} that underlies the Poisson process.

While it is, as a matter of principle, necessary to account for the effect of the detector in \emph{any} experiment, the extent to which the physics of the detector needs to be taken seriously for predicting arrival times is not self-evident. In practice, scattering experiments such as the double-slit and the Stern-Gerlach experiment are routinely analyzed with no reference whatsoever to the detector. But since the detectors employed in these experiments are typically no more specialized than the ones found in TOF experiments (e.g., a scintillation screen employed in \cite{Pfau,*wig,*Pfaudetector}), it is not a priori obvious why the physics of the detector is any more relevant for predicting the statistics of arrival times than it is for predicting the statistics of impact positions. Hence it should not come as a surprise that the quantum flux distribution gives the expected result despite ignoring the detector. It has even been shown that \(\Pi_{\text{QF}}(\tau)\) can arise from a careful consideration of a physical detector, e.g., a laser curtain inducing fluorescence from an incoming atom \cite{Mugaphoton,MugaReal,3DMuga}. This suggests that one should turn to realistic TOF experiments if one wants to take the detector seriously.

Finally, the semiclassical distribution depicted in Fig.\ \ref{SCQF} also fails to display the expected behavior since it is largely supported around \(\tau_0\). This distribution is often used in the experimental determination of the momentum distribution. Using the measured arrival-time distribution \(\Pi_{\text{meas}}(\tau)\), the empirical momentum distribution is taken to be
\begin{equation}
    \frac{mL}{p^2}\sm\Pi_{\text{meas}}\!\left(\frac{mL}{p}\right)\kern-0.1em,
\end{equation}
corresponding to the quantum mechanical momentum distribution \(\hbar^{-1}\sm|\tpsi(p\smalldiv\hbar)|^2\), thereby tacitly assuming the validity of \eqref{Pisc}. However, our results indicate that such reconstructions are questionable (see also \cite[Ch.\ 4]{NicolaPhd}). 

\section*{Acknowledgements}
W.S.\ is supported by the Research Foundation Flanders (Fonds Wetenschappelijk Onderzoek, FWO), Grant No.\ G066918N. It is a pleasure to thank J.\ M.\ Wilke for valuable editorial input.

\bibliography{ref}
\end{document}